\documentclass[11pt,letterpaper]{article}
\usepackage[utf8]{inputenc}
\usepackage{amsmath,amsfonts}
\usepackage{algorithmic}
\usepackage{algorithm}
\usepackage{array}
\usepackage[caption=false,font=normalsize,labelfont=sf,textfont=sf]{subfig}
\usepackage{textcomp}
\usepackage{stfloats}
\usepackage{url}
\usepackage{verbatim}
\usepackage{graphicx}
\usepackage{hyperref}
\usepackage{cite}
\usepackage{algorithm, algorithmic, amssymb, bbm, bm}
\DeclareMathOperator*{\argmax}{arg\,max}

\usepackage[
letterpaper,
top=1in,
bottom=1in,
left=1in,
right=1in]{geometry}

\usepackage[x11names]{xcolor}
\hypersetup{
  colorlinks=true,
  citecolor=SpringGreen4
}

\begin{document}

\title{Quantum-Inspired Approximations to \\ Constraint Satisfaction Problems}
\date{}
\author{S. Andrew Lanham  
    \thanks{Applied Research Laboratories, The University of Texas at Austin, Austin, TX 78758 USA (e-mail: sa\_lanham@utexas.edu).}}

\maketitle

\begin{abstract}
Two contrasting algorithmic paradigms for constraint satisfaction problems are successive local explorations of neighboring configurations versus producing new configurations using global information about the problem (e.g. approximating the marginals of the probability distribution which is uniform over satisfying configurations). This paper presents new algorithms for the latter framework, ultimately producing estimates for satisfying configurations using methods from Boolean Fourier analysis. The approach is broadly inspired by the quantum amplitude amplification algorithm in that it maximally increases the amplitude of the approximation function over satisfying configurations given sequential refinements. We demonstrate that satisfying solutions may be retrieved in a process analogous to quantum measurement made efficient by sparsity in the Fourier domain, and present a complete solver construction using this novel approximation. Freedom in the refinement strategy invites further opportunities to design solvers in an evolutionary computing framework. Results demonstrate competitive performance against local solvers for the Boolean satisfiability (SAT) problem, encouraging future work in understanding the connections between Boolean Fourier analysis and constraint satisfaction.
\end{abstract}

\section{Introduction}
\label{sec:introduction}

Constraint satisfaction (CS) problems are a central, ubiquitous class of problems in theoretical and applied computer science. Algorithms for solving these problems find wide application in optimization and also shed light on the broader theory of computational hardness. It is useful to break down solvers for CS into two broad categories. The first category encompasses \textit{local solvers} while we refer to the second category as \textit{global solvers}. In the local paradigm, a particular configuration of variables is selected and the constraints are evaluated at that point to verify satisfiability. If not all of the constraints have been satisfied, a new point is generated to explore sequentially, typically based on the remaining unsatisfied constraints, or by backtracking to a previous configuration and branching (see, e.g., \cite{davis1962machine,spears1993simulated,selman1993local, glover1998tabu}). 

For designing global solvers, on the other hand, a typical motivating thought experiment is to consider a hypothetical ``oracle" returning samples from the uniform distribution over satisfying assignments of a given CS formula \cite{altarelli2009connections}. We refer to this distribution as the ``oracle" distribution. We can frame the global solver design approach as one which produces sequential estimates that attempt to capture samples from this particular distribution, but using non-local methods. In global evolutionary approaches, for example, candidate solutions may be generated using combinations of previous solutions rather than direct neighbors of a single previous solution, representing a type of ``globally-aware" local approach \cite{qasem2009learning,folino2001parallel,prugel2011maximum}. Other global approaches use mappings to different problems, e.g. coloring, to derive solvers \cite{flaxman2003spectral}. A separate class of global approaches focuses on forming direct approximations of the marginal distributions of the oracle distribution, in particular, the marginals corresponding to each bit of the satisfying Boolean assignment. Message passing (MP) procedures, for example, have been used in the CS paradigm to produce these estimates, effectively using belief propagation (BP) to solve problems \cite{hsu2006characterizing,braunstein2005survey}. While  BP methods seem to perform best for randomly-generated SAT instances \cite{hsu2006characterizing}, there exist very few alternative methods that approximate marginal distributions directly. Our work targets this second class of global solvers and, in particular, presents approximations for the distribution over satisfying solutions to a CS problem. 

The algorithms we propose differ from the methods above, however, in the approximation strategy. The framework adopted here is  ``quantum-inspired," and we demonstrate the relevance of quantum information processing techniques to classical CS solvers.  Quantum-inspired classical computing has previously yielded competitive numerical techniques in linear algebra, machine learning, and genetic algorithms \cite{tang2019quantum, shao2022faster, narayanan1996quantum}, suggesting a broad applicability of the methods of quantum information processing. The Boolean satisfiability problem, in particular, serves as a productive area to study the interrelation between classical and quantum algorithms. CS is addressed in the quantum realm primarily by Grover's algorithm, which provides a fundamental improvement in worst-case query complexity and is generalizable via \textit{amplitude amplification} to a wide array of problems \cite{grover1996fast, brassard2002quantum}. The quantum approaches differ significantly in structure from canonical non-quantum CS solvers \cite{biere2009handbook,marques1999grasp,moskewicz2001chaff}. The success of nature-inspired and evolutionary computing algorithms in this area, however, (e.g. the relevance of cavity methods of statistical physics in characterizing the difficulty of SAT problems \cite{mezard2002analytic,braunstein2004survey} or the use of genetic algorithms \cite{folino2001parallel,dozier1998solving,craenen2003comparing}) motivates investigation of more directly quantum-inspired classical computing techniques, an underexplored area for CS and SAT.

Grover's algorithm solves the Boolean SAT problem by iteratively evolving a quantum state in an uninformative superposition to a new state which is maximally separated from the initial superposition state in $\ell_2$ distance over the satisfying solutions \cite{bennett1997strengths}. In other words, the probability of measuring satisfying solutions in the final state is as high as possible with respect to the initial state. Our quantum-inspired approximation procedure can be viewed as one of constructing an iteratively improving but initially low-complexity approximation to the probability distribution with the statistical properties of the final state produced by Grover's algorithm. Furthermore, the methods we propose for accessing satisfiable solutions from this approximation are directly inspired by the qubit measurement paradigm (although an $\ell_2$ measurement framework is not a hard requirement in a classical setting as it is for quantum). In our approximation algorithm, the intermediate steps for computing the approximation are not quantum-inspired, however, in the sense that we do not adopt unnecessary restrictions like unitary evolution. 

Our novel paradigm for solving CS problems instead relies mainly on concepts from Fourier analysis of Boolean functions and uses them to mirror the transformations characteristic of amplitude amplification, but at the input-output level. In particular, the approach leverages relationships between Boolean indicator function representations of individual constraints to build efficient approximations. It is dependent on two crucial properties that are features of many CS problems. The first feature is the modularity of the constraints in the problem description: the principle that if the entire formula is satisfied for some configuration, then any subset of constraints is also satisfied for the same configuration. The second feature is the existence of a factorized representation of the constraint satisfaction formula, where each constraint is individually representable as a relatively sparse function in the Fourier domain. These two properties are not universal to all constraint satisfaction problems, but all problems with these properties are amenable to our approximation method. Incidentally, these properties are also what allow for practical oracle operator design in the quantum algorithms framework, and so it is not surprising that the same principles arise in our iterative approximation framework. 

The main focus in this work is on creating novel solvers whose performance we evaluate in Section \ref{sec:solver} in non speed-optimized settings. In the framework of global solvers, our design approach is a method for iteratively \textit{learning} the marginals of a complicated ``oracle" distribution using low-degree polynomials. This invites comparisons to the broad area of learning Boolean functions, a rich subject relying heavily on methods of Boolean Fourier analysis \cite{o2014analysis, linial1993constant, kushilevitz1993learning}. The work also relates to the general framework of the polynomial method as applied to algorithm design \cite{williams2014polynomial,abboud2014more}, where a difficult computational task is simplified through the use of low-degree polynomials and associated algorithmic methods. Construction methods for low-degree approximations to pseudo-Boolean functions arose out of investigations of Boolean switching functions, and some best-linear and $\ell_2$ approximations have been characterized in detail \cite{hammer1992approximations, boros2002pseudo, coleman1961orthogonal, o2014analysis}. The performance of low-degree polynomial approximation methods was characterized extensively in relation to SAT in \cite{bresler2022algorithmic}. Other approaches to SAT based on Boolean Fourier analysis have yielded competitive local solvers \cite{kyrillidis2020fouriersat}. Similarly, there are low-degree methods for SAT which fall under the ``global" paradigm \cite{williams2014polynomial}. We note that techniques involving low-degree Boolean polynomials are also ubiquitous in studies of quantum query complexity \cite{beals2001quantum,aaronson2008polynomial}, and in quantum algorithm design.

Our solver approach iteratively refines a simple approximation to the oracle distribution based on information about unsatisfying solutions and violated constraints. It differs especially from previous approximation methods by the manner in which information is accessed; in this case, a quantum measurement-inspired framework. Furthermore, there is a combinatorial explosion in the number of possible refinements to compute, and so parsimonious refinement strategies are required to manage the complexity of the approximation. The freedom of choice in the computation path also invites an analysis of methods for intelligently navigating the tradeoff between efficient and useful refinements. We demonstrate, however, that even random refinements can greatly improve the success rate of the final CS solver. Freedom of selection in the refinement strategy invites consideration of evolutionary approaches, such as intelligent branching into child refinements and optimal metrics for improving a refinement. The results in this area can guide future analysis on the targeted learnability of high-degree functions and the resulting accuracy of marginal estimates, further bridging the concepts of satisfiability solvers and learnability.

\begin{figure}[ht!]
    \centering
    \includegraphics[width=\linewidth]{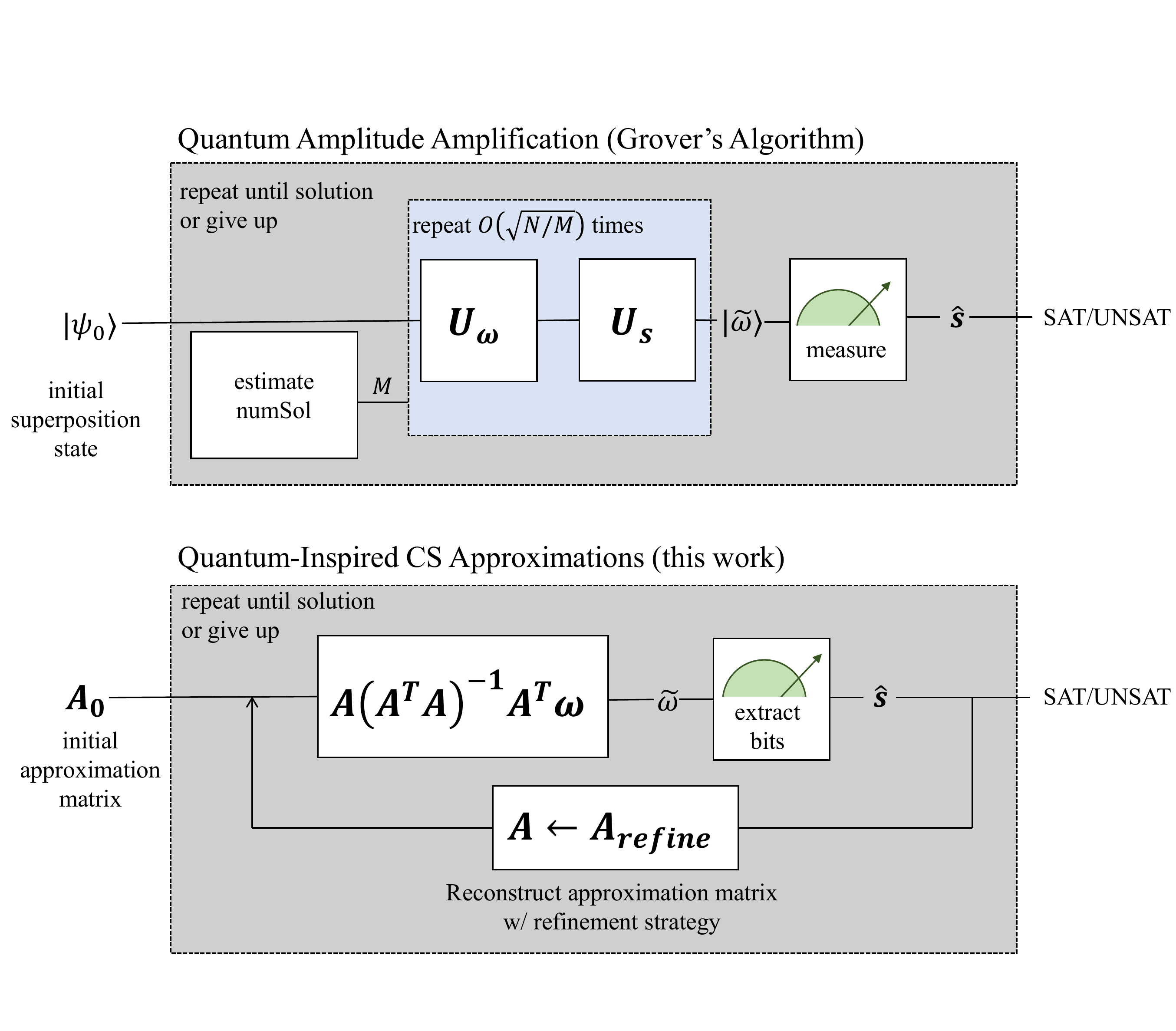}
    \caption{The high-level processing diagrams for amplitude amplification and our novel CS approximation framework highlights the quantum-inspired design approach. Quantum amplitude amplification requires either foreknowledge of the number of marked items (solution), or else an estimate must be given, since this fixes the number of iterations of the algorithm. An example estimation procedure is detailed in \cite{brassard2002quantum}. Quantum states are additionally restricted to unitary evolution; absent such a requirement, our CS approximation algorithm uses significantly different intermediate processing methods.}
    \label{fig:block_diagram}
\end{figure}

We continue in Section \ref{sec:prelim} with some basic results of Boolean Fourier analysis and examine the modularity and factorization properties highlighted above in more detail. In Section \ref{sec:algo} we present the main approximation algorithm, while in Section \ref{sec:solver} we leverage this approximation to form a CS solver and characterize its performance. We focus our solver performance characterization using  Boolean $k$-conjunctive normal form ($k$-CNF) SAT problems as benchmarks due to their ubiquity and the fact that the techniques can be generalized easily to broader CS problems. We conclude with a brief discussion of major tradeoffs arising in the solver design and applicability to more general evolutionary computing paradigms.

\section{Preliminaries}
\label{sec:prelim}
We review the relevant machinery for analysis of Boolean functions \cite{o2014analysis}. A pseudo-Boolean function is a mapping from an $n$-bit binary string taking values $-1$ or $1$ to a real number, i.e. $f(\mathbf{s}) : \{ -1, 1\}^n \mapsto \mathbb{R}$. Let $\mathbf{f}\in\mathbb{R}^{2^{n}}$ be a vector containing an enumeration of $f(\mathbf{s})$ over all possible inputs. In the vector representation, the position of each entry corresponds to an input $\mathbf{s} \in \{-1, 1\}^n$ using standard binary ordering, while the value at the position is the value $f(\mathbf{s})$. Any pseudo-Boolean function of the given form can be expressed as a multilinear polynomial, constituting the \textit{Fourier expansion} of $f(\mathbf{s})$ \cite{o2014analysis}. Letting $[n] = \{0, \ldots, n-1\}$, this expansion is given as $f(\mathbf{s}) = \sum_{S \subseteq [n]} \hat{f}(S)x^S $ where the monomials $x^S$  are defined as $x^S = \prod_{i \in S} x_i$ with $x^{\emptyset} = 1$ and  the coefficients $\hat{f}(S)$ form the \textit{spectrum} of $f(\mathbf{s})$. By Plancherel's Theorem, we have  for any functions $f$, $g : \{-1, 1\}^n \mapsto \mathbb{R}$ that 
\begin{equation}
    \frac{1}{2^n} \sum_{x \in \{-1, 1\}^n} f(x) g(x) = \sum_{S \subseteq [n]} \hat{f}(S) \hat{g}(S) \, .
\end{equation}

We now  re-express SAT problems in the framework of pseudo-Boolean analysis. We consider SAT propositional formulas expressed in the conjunctive normal form. In CNF, formulas are expressed as a series of clauses joined by conjunction (logical ``and" operation). A $k$-CNF SAT \textit{clause} is a disjunction (logical ``or" operation) between $k$ literals, where a literal is a Boolean variable $x_i$ or its complement $\bar{x}_i$. As an example, a 3-SAT clause could take the form $C_m = (x_1 \lor x_4 \lor \bar{x}_7 )$. Let $F$ be a $k$-SAT propositional formula of $M$ clauses in $\{0,1\}$ representation; then we have $ F(\mathbf{x}) = \bigwedge_{m = 1}^{M} C_m$. Each clause comprising a propositional formula can now be mapped to a pseudo-Boolean indicator function, which conventionally returns $1$ if the clause is satisfied for an input argument in $\{-1, 1\}^n$ and $0$ if it is not. 

For our final solver construction, however, it will be more important to define the indicator function of the clause \textit{complement}, i.e., the function that returns $1$ where the clause is \textit{not} satisfied and zero elsewhere. We choose a widely adopted multilinear expansion for the SAT clause complement indicator function \cite{gu1999optimizing, ercsey2011optimization}, where for clause $m$ we define 
\begin{equation}
    \label{eq:inv_ind}
    k_m(\mathbf{s}) \triangleq \frac{1}{2^k}\prod_{i = 1}^{n} (1-c_i s_i) \, ,
\end{equation}
with $ c_i = -1$ if $\bar{x_i}$ appears in clause $C_m$, $c_i = 1$ if $x_i$ appears in $C_m$, and $c_i = 0$ if neither $x_i$ nor $\bar{x_i}$ appear. Because all subsequent analysis uses the indicator functions of Equation (\ref{eq:inv_ind}), we simply refer to them as ``the (clause) indicator functions" with the understanding that the typical ``on"/``off" convention is reversed. It is straightforward to see that the largest multilinear expansion of a single $k$-CNF SAT clause indicator will contain $2^k$ unique terms. In this sense, the multilinear representation of each $k_m(\mathbf{s})$ serves as a sparse alternative to the full enumeration over all inputs, $\mathbf{k}_m$, since the latter has length $2^n$ while the former, a Fourier expansion, captures the clause constraint using $\mathcal{O}(2^k)$ coefficients.

We also note that the hypothetical ``oracle distribution" introduced in Section \ref{sec:introduction} can be constructed using only the indicator functions of Equation (\ref{eq:inv_ind}), as long as it is properly divided by a partition function. Let $\omega_F(\mathbf{s})$ be the function which is ``on" for inputs which are solutions to a satisfiability formula $F$. We refer to $\omega_F(\mathbf{s})$  as an ``\textit{oracle function}" and drop the subscript $F$ henceforth, understanding that the oracle is problem-specific. By definition, we have 
\begin{equation}
    \omega(\mathbf{s}) \triangleq \prod_{m=1}^{M} (1 - k_m(\mathbf{s}) )
\end{equation}
where index $m$ iterates over each clause in the formula. With normalization by the number of satisfying solutions, the function $\omega(\mathbf{s})$ is the probability mass function of the uniform distribution over solutions. The ensuing sections focus on building estimates related to partitions of $\omega(\mathbf{s})$ induced by bits $s_i$.

\subsection{Key Properties of Indicator Functions}
The function $\omega(\mathbf{s})$ exhibits the two crucial features highlighted in Section \ref{sec:introduction} -- modularity, and representation with sparse factors. The modularity property refers to the fact that each individual factor  $(1-k_m(\mathbf{s}))$ constituting $\omega(\mathbf{s})$ must be individually satisfied when $\mathbf{s}$ is a satisfying solution, i.e., for satisfying configuration $\mathbf{s}^*$ we have 
\begin{equation}
    \label{eq:modular}
    \prod_{\ell \in S} (1-k_{\ell}(\mathbf{s}^*)) = 1 \; \quad \forall S  \subseteq [M] \,.
\end{equation}
These relations immediately follow from $\omega(\mathbf{s}^*) = 1$ and the fact that the indicator functions take only values on $\{0,1\}$. Furthermore, the factorized representation of $\omega(\mathbf{s})$ with factors of $\mathcal{O}(2^k)$ coefficients each allows for certain modular relations arising from Equation (\ref{eq:modular}) to be expressed as constraints on functions of low degree and low relative complexity. This property is useful in reducing the computational overhead of the algorithm of Section \ref{sec:algo} both in the computation of the approximation and the extraction of candidate solutions. 

\subsection{Measuring Partitions of Boolean Functions}
We briefly define some useful quantities which are proportional to the partitions of a Boolean function. We later demonstrate that through Plancherel's Theorem, these quantities can be  efficiently accessed using Fourier coefficients. 
Let $s_i$ be the $i^{\textrm{th}}$ bit of some input $\mathbf{s} \in \{-1,1\}^n$.
Given a pseudo-Boolean function $f(\mathbf{s})$ and $b \in \{-1, 1\}$, we can define the sum $\sum_{\mathbf{s}} f(\mathbf{s} | s_i = b) = \sum_{\mathbf{s} : s_i = b} f(\mathbf{s})$. With this notation, we introduce a measure of the \textit{bias} of a bit relative to a function. We give an example measure of bias capturing a bit-partitioning:
\begin{equation}
\label{eq:bias1}
\beta_i(f) \triangleq \sum_{\mathbf{s}} f(\mathbf{s} | s_i = 1) - f(\mathbf{s} | s_i = -1) \, . 
\end{equation}  

The biases $\beta_i(\cdot)$ can be also expressed as inner products. Consider the single-bit functions $e_i(\mathbf{s}) = s_i$, noting that they take value $1$ when bit $i$ of the input $\mathbf{s}$ is $1$ and $-1$ otherwise. We have for general function $f(\mathbf{s})$ that $\langle \mathbf{e}_i , \mathbf{f} \rangle = \sum_{\mathbf{s}} f(\mathbf{s} | s_i = 1) - \sum_{\mathbf{s} } f(\mathbf{s} |  s_i = -1)$. Using Plancherel's theorem, we can equivalently say that $\langle \mathbf{e}_i , \mathbf{f} \rangle = 2^n \sum_{S \subseteq [n]} \hat{e}_i(S) \hat{f}(S)$, i.e. that
\begin{equation} 
\sum_{\mathbf{s} } f(\mathbf{s} | s_i = 1) -  f(\mathbf{s} | s_i = -1) = 2^n \hat{f}(s_i) 
\end{equation}
This indicates that with just one coefficient of the Fourier expansion of  $f(\mathbf{s})$, we may deduce a measure of bias related to the weight over a particular bit, i.e. whether more function mass is concentrated at $s_i = 1$ or $-1$. Related quantities arise usefully in inference problems such as maximum likelihood estimation over marginal distributions, highlighting the relevance of Fourier analytic techniques. 

Other measures of bias are possible. For example, Parseval's Theorem provides an equivalence $\frac{1}{2^n} \sum_{\mathbf{s}} f(\mathbf{s})^2 = \sum_{S \subseteq [n]} \hat{f}(S)^2$, which allows for computation of the function $\ell_2$ norms in the Fourier domain. The difference in $\ell_2$ norms between function partitions yields a useful measure of bias, i.e., 
\begin{equation}
    \label{eq:ell2bias}
    B_{2,i}(f) \triangleq || f(\mathbf{s} | s_i = 1 )||_2^2 - || f(\mathbf{s} | s_i = -1 )||_2^2 \, ,
\end{equation} and, by Parseval's Theorem, we have a potentially simplified computation, $B_{2,i}(f) = \sum_{S} \hat{f}(S | s_i = 1)^2 - \hat{f}(S | s_i = -1)^2$. These quantities draw analogy to the Born rule statistics, which are based on the same individual norms. Efficient computation is facilitated if there are a sparse number of Fourier coefficients relative to the enumerated vector $\mathbf{f}$.

\section{Approximation Algorithm}
\label{sec:algo}
In applying these concepts to SAT solvers, we broadly mirror the paradigm of quantum algorithms by designing a procedure that creates an iteratively improving approximation to the oracle function $\omega(\mathbf{s})$. The measured biases of the approximating function evolve to produce satisfying configurations. In our more flexible classical paradigm, we may consider the measures of bias characterized in the previous section, i.e. Equations (\ref{eq:bias1}) and (\ref{eq:ell2bias}). These types of quantities are deterministic as opposed to the constrained quantum setting requiring the inherently random Born rule. The approach is fundamentally similar to other quantum-inspired design paradigms by substituting straightforward querying of individual points with sampling-relevant summations (here arising from the Fourier coefficients; see, e.g. \cite{tang2019quantum}).

Our first-order approach provides a method for approximating the oracle function $\omega(\mathbf{s})$ using weighted combinations of the functions $k_m(\mathbf{s})$ for all $m \in [M]$ followed by a method for calculating the bit-biases of the final approximation. Each clause constraint encodes information about the solution space $\omega(\mathbf{s})$ which is highlighted by the modularity property of Equation (\ref{eq:modular}). Furthermore, the $\mathcal{O}(2^k)$-length Fourier description of each $k_m(\mathbf{s})$  facilitates efficient computation of relevant biases. The proposed method can be easily generalized to use second and higher-order products of indicator functions $k_m(\mathbf{s}) k_n(\mathbf{s}), \, m \neq n$ or higher. 

Given a family of $M$ clause indicators representing SAT formula constraints, we now detail a first-order $\ell_2$-approximation for $\omega(\mathbf{s})$ using these indicators. In vector notation, the best approximation of the vector $\bm{\omega}$ within the column space of the matrix $\mathbf{A}$ is expressed as 
\begin{equation}
    \label{eq:approximation}
    \tilde{\bm{\omega}} = \mathbf{A} (\mathbf{A}^T \mathbf{A})^{-1} \mathbf{A}^T \bm{\omega} \, .
\end{equation}
When the columns of $\mathbf{A}$ are taken to be clause indicators $\mathbf{k}_m$, the best approximation may be computed using quantities efficiently derived from Fourier coefficients. 

Recall that a $k$-CNF-SAT problem is parametrized by $n$ variables and $M$ clauses. For the first-order approximation, we consider a $2^n \times (M+1)$  approximating matrix $\mathbf{A}$, which we describe in terms of its columns. Let the first column of $\mathbf{A}$ be a column of all $1$s. Let each of the $M$ subsequent columns of $\mathbf{A}$ contain the vector enumerations of indicators $k_m(\mathbf{s})$, $\mathbf{k}_m$. We demonstrate that all relevant computations for Equation (\ref{eq:approximation}) are polynomial in $M$ and $2^k$, where $M$ is typically $\mathcal{O}(n)$ and $k$ is constant in $n$. Consider first the rightmost multiplication, $\mathbf{A}^T \bm{\omega}$. We adopt the informal heuristic that $[\mathbf{A}]_0^T \bm{\omega} = \mathbbm{1}^T \bm{\omega} \triangleq 1$, i.e. oracle vector $\bm{\omega}$ has a nonzero overlap with the vector of all ones (a solution exists). In the context of a SAT solver, the exact value of the inner product turns out to be immaterial because (1): the final solver is \textit{incomplete} and (2): the bit-biases are insensitive to scale. Incomplete solvers are those which do not terminate unless they find a solution (or a timeout period is reached \cite{kautz2009incomplete}). The inner product reflects an assumption that a solution exists, and so it will have positive overlap with a constant function; if a solution does not actually exist, then incomplete solvers will simply terminate after a given timeout is reached.

We secondly observe that inner products between $\bm{\omega}$ and any $\mathbf{k}_m$ are zero, 
\begin{equation} 
\langle \mathbf{k}_m ,\bm{\omega} \rangle = 0 \quad \forall m \in [M] \, .
\end{equation}
This crucial fact follows from the modularity property of Equation (\ref{eq:modular}) and the construction of the indicator function in Equation (\ref{eq:inv_ind}), taking value $1$ only where clause $m$ is not satisfied. Put another way, the vectors are orthogonal because solutions must satisfy all clauses, and $\bm{\omega}$ is only nonzero at solutions. From these two observations, we conclude that 
\begin{equation} 
\mathbf{A}^T \bm{\omega} = [1, 0, \ldots , 0]^T \, ,
\end{equation}
allowing precomputation of the rightmost multiplication of Equation (\ref{eq:approximation}) for any problem. 

Turning to the computation of $( \mathbf{A}^T \mathbf{A})^{-1}$, an $(M+1)\times (M+1)$ matrix, we recall that the indicator functions of Equation (\ref{eq:inv_ind}) comprising the columns of $\mathbf{A}$ are $2^k$-sparse in their Fourier representation, while the $\mathbbm{1}$ vector is 1-sparse. By Plancherel's Theorem, the quantities $\langle \mathbf{k}_m, \mathbf{k}_n \rangle = \sum_{S \subseteq [n]} \hat{k}_m(S) \hat{k}_n(S) $ are computable in $\mathcal{O}(2^{2k})$ operations. Since $[\mathbf{A}^T \mathbf{A}]_{i,j} = \langle \mathbf{k}_i ,  \mathbf{k}_j \rangle$, we can construct $\mathbf{A}^T \mathbf{A}$ in $\mathcal{O}(2^{2k} M^2)$ operations and invert it in $\mathcal{O}(M^3)$ operations using worst-case matrix-inversion results. The process yields a reduced-complexity computation of the $M+1$ coefficients arising from $(\mathbf{A}^T\mathbf{A})^{-1}\mathbf{A}^T  \bm{\omega}$ without requiring full enumeration of $\mathbf{A}$ and $\bm{\omega}$. The best approximation is then $\hat{\bm{\omega}} = \sum_i a_i [\mathbf{A}]_i$, where $[\mathbf{A}]_i$ is the $i^{\mathrm{th}}$ column of $\mathbf{A}$. Equivalently, 
\begin{equation}
    \label{eq:best_approx}
    \hat{\bm{\omega}} = a_0 \mathbbm{1} + \sum_{j=1}^{M} a_j \mathbf{k}_{j-1}
 \, ,
\end{equation} with $ a_i = [(\mathbf{A}^T\mathbf{A})^{-1}\mathbf{A}^T  \bm{\omega}]_i$. The coefficients $a_i$ form the weights for our linear best-approximation of $\bm{\omega}$ using only the indicators for individual $k$-CNF SAT clause constraints.

The approximation method is generalized to second-order by forming and including the products between Fourier coefficients of pairs of indicator vectors, e.g., $\hat{k}_m(S) \hat{k}_n(S)$. This corresponds to  pointwise (Hadamard) products $\mathbf{k}_m \odot \mathbf{k}_n$ and can be recursed to higher order. Products of indicator functions will also have zero inner product with $\bm{\omega}$, just as first-order constructions, by Equation (\ref{eq:modular}). Our SAT solver construction in Section \ref{sec:solver} uses up to second-order indicator functions, and performance generally improves with more indicators, but with diminishing returns as both computational overhead grows and the products of indicators reduce the domain of the  resulting function which is nonzero. The computation complexity increases exponentially with the order of the product; multiplication of $v$ clause indicators requires $\mathcal{O}(2^{kv})$ multiplications of Fourier coefficients.

In standard quantum amplitude amplification, the final state perfectly encodes $\bm{\omega}$. Our approximation process can be seen as a crude classical analog to amplitude amplification in that it builds a function with maximally overlapping inner product with $\bm{\omega}$, given $\mathbf{A}$. In practice, however, the columns of $\mathbf{A}$ are low-order clause indicators, yielding a low-fidelity approximation. We claim that classical access methods for computing bit biases partially offset the inaccuracies, however. It is an interesting question to explore how quickly our approximation converges to $\bm{\omega}$ as the order of the approximation increases. To this end, CS models for planted solutions may serve as useful analytical tools \cite{feldman2015complexity}. 

\begin{algorithm}
\caption{AmplificationSAT Solver}
\begin{algorithmic}[H!]
 \renewcommand{\algorithmicrequire}{\textbf{Input:} $M$ $k$-CNF clauses, timeout $T_{\mathrm{timeout}}$} 
 \renewcommand{\algorithmicensure}{\textbf{Output:} candidate solution $\mathbf{s}_{\mathrm{final}}$}
 \REQUIRE in
 \ENSURE  out \\ 
  \textit{Initialization}: $r \leftarrow 1, \; t_{\mathrm{elapsed}} \leftarrow 0 $ \\ 
  \STATE compute first-order approximation in Fourier domain: \\  
  \textrm{          } $\tilde{\bm{\omega}} = \mathbf{A} (\mathbf{A}^T  \mathbf{A})^{-1} \mathbf{A}^T \bm{\omega}$ 
  \STATE compute $\mathbf{s}_* = $  MeasureBias($\tilde{\bm{\omega}}$) to form solution estimate  \\ 
  \STATE set $\mathbf{A}_{\mathrm{prev}} \leftarrow \mathbf{A}$
  \WHILE {$\mathbf{s}_*$ is not satisfying AND $t_{\mathrm{elapsed}} < T_{\mathrm{timeout}}$} 
  \STATE gather clause indices $U = \textrm{ClauseNeighbors}(\mathbf{s}_*)$. \\  
  \IF {all pairs from $U$ are already columns of $\mathbf{A}_{\mathrm{prev}}$} 
  \STATE pick random clause index $p = \mathrm{rand}\{0, \ldots, M-1\}$
  \STATE form Fourier coeffs of all pairs $\mathbf{k}_p \odot \mathbf{k}_j, \, j \in [M]$
  \ELSE {
  \STATE form Fourier coeffs of all pairs $\mathbf{k}_i \odot \mathbf{k}_j, \, i, j \in U$ }
  \ENDIF
  \STATE treat new pairs as new columns; forms $\mathbf{A}_{\mathrm{new}}$ \\ 
  \STATE compute new approximation $\tilde{\bm{\omega}}$ with $\mathbf{A}_{\mathrm{new}}$  
  \STATE compute new $\mathbf{s}_* = \textrm{MeasureBias}(\tilde{\bm{\omega}})$ 
  \STATE $\mathbf{A}_{\mathrm{prev}} \leftarrow \mathbf{A}_{\mathrm{new}}$ 
  \STATE $r \leftarrow r + 1$
  \STATE $\mathbf{s}_{\textrm{final}} = \textrm{LocalSearch}(\mathbf{s}_*)$
  \ENDWHILE
 \RETURN $\mathbf{s}_{\textrm{final}}$ 
 \end{algorithmic} 
\end{algorithm} 

\section{Candidate Solver and Performance}
\label{sec:solver}

\begin{figure}[ht!]
\begin{center}
\includegraphics[width=\linewidth]{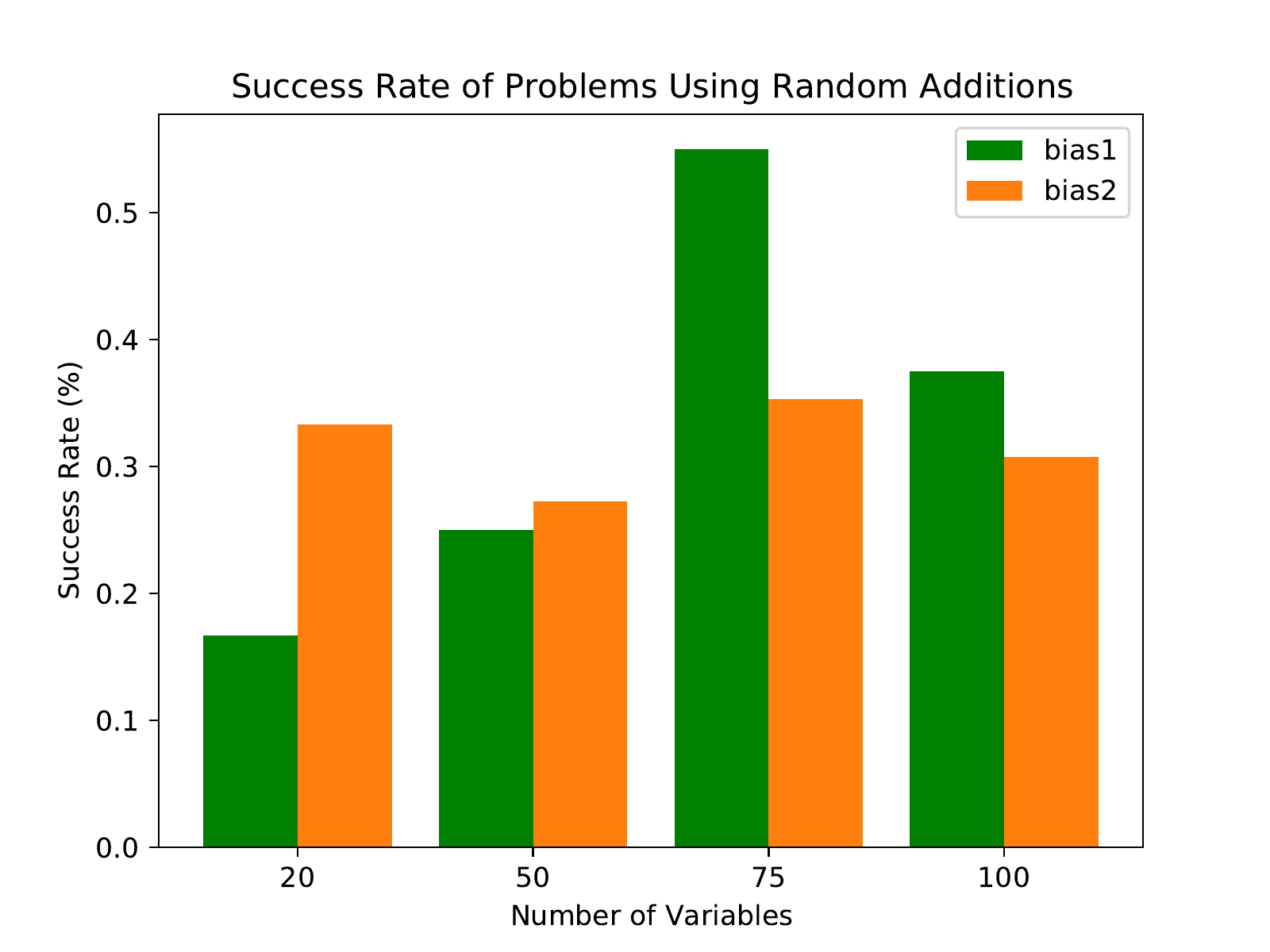}
\end{center}
\caption{The chart demonstrates the utility of random approximation refinements on the success rate by depicting the success rate on only the instances which required random refinements (the criteria for which is detailed in Algorithm 1). A significant percentage of problems which were not solvable using ClauseNeighbors were later solved with the addition of random approximation refinements.}
\label{fig:randomness_success}
\end{figure}

In this section we present a full SAT solver based on the approximation of Section \ref{sec:algo} and assess its performance on benchmark problems against other solvers. The solver approach begins by computing an approximation $\tilde{\bm{\omega}}$ to $\bm{\omega}$. The biases associated with each bit-partition are obtained using a simple decimation procedure, with choice of bias function. This produces a candidate solution. If the solution is not satisfying, a heuristic algorithm is used to refine the approximation based on remaining unsatisfying clauses, and the approximation is re-computed with relevant higher-order indicator functions in the column space of $\mathbf{A}$. This process is repeated until either a satisfying solution is found or the process times out. Sometimes it is prohibitive to refine the approximation using heuristics because all relevant heuristic refinements have been already integrated into the approximation. In these cases the solver selects a random clause and adds all second-order indicator pairs that include that clause. While the heuristic motivation is less obvious for the random approach, the results indicate that the introduction of randomness helps break approximations out of unsatisfying clusters (Figure \ref{fig:randomness_success}). Note also that after each round of approximation, the solver proceeds to a local search (in our case, low temperature annealing) proceeding from the output point of the approximation. This allows exploration of the neighborhood around each approximation and singles out approximations which are close in Hamming space to satisfying solutions. If the solver fails to find a solution before the timeout period is exceeded, the formula is declared unsatisfiable. 

 \begin{figure*}[]
\begin{center}
\includegraphics[width=\textwidth]{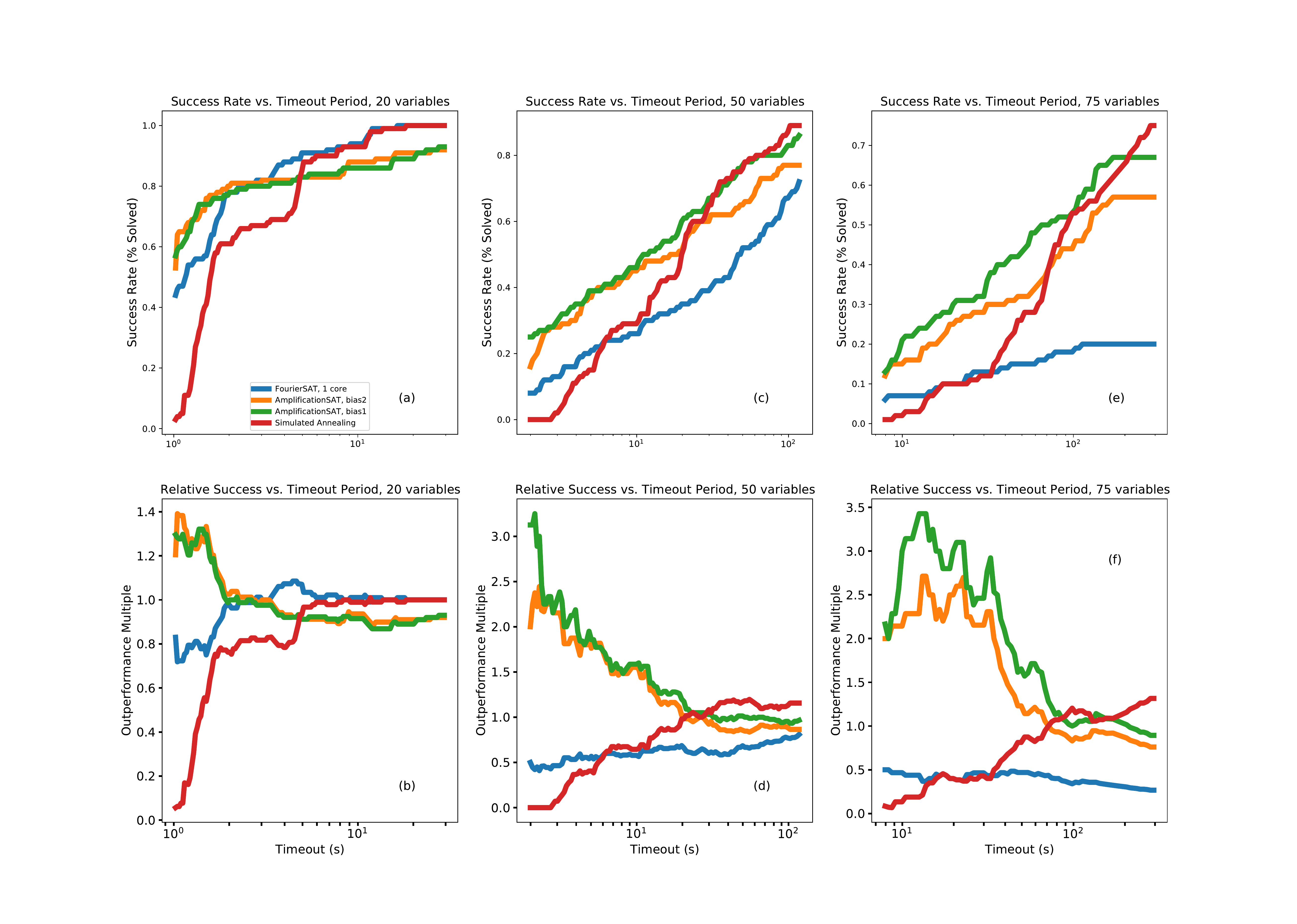}
\end{center}
\caption{Performance over timeouts ranging from 2s to a variable-dependent upper limit (60s, 120s, and 180s respectively). Outperformance multiple is defined as the multiple of problems solved by a solver relative to the best solver of the other two at any given timeout. AmplificationSAT has a higher success rate on short timescales whereas local solvers are hampered by randomized start points and subsequent burn-in times. This suggests the utility of an AmplificationSAT result as a tool for generating a rapid heuristic guess at a solution. For 20 variable problems, a 1000 step annealing schedule was chosen, while for 50 and 75 variable problems the steps were increased to 2000 and 4000 respectively. Temperature parameters are defined in Section \ref{sec:res_num}}
\label{fig:adv_comp}
\end{figure*}
 
  \begin{algorithm}
 \caption{MeasureBias}
 \begin{algorithmic}
 \renewcommand{\algorithmicrequire}{\textbf{Input:} Approximation function $f$}
 \renewcommand{\algorithmicensure}{\textbf{Output:} Solution $\mathbf{s}_*$} 
 \REQUIRE in 
 \ENSURE out \\
 \textit{Initialization}: $V \leftarrow [n]$
 \FOR{$i = 1$ to $n$}
    \STATE $B_{i_*} = \max_{i \in V} \hat{f}(s_i)$ 
    \IF{  $B_{i_*} > 0$}
    \STATE $s_{i_*} = 1$   
    \ELSE 
    \STATE $s_{i_*} = -1$
    \ENDIF
    \STATE $\hat{f}(S) \leftarrow \hat{f}(S | s_i = s_{i_*}) $
    \STATE $V \leftarrow V \setminus \{ i_* \}$
 \ENDFOR
 \STATE \textbf{return} $\mathbf{s}_*$
 \end{algorithmic}
\end{algorithm}

 We first introduce the operation $\textrm{MeasureBias}(\tilde{f})$ used by our solver for obtaining candidate solutions from the approximation. Let $B_i(\cdot)$ be a general measure of bias for bit $i$. MeasureBias computes the quantities $i_* = \argmax_{i \in [n] \setminus V} |B_i(\tilde{f})|$ and $B_* =  B_{i_*}(\tilde{f})$, where $V$ is initially empty, but holds previously selected variables as the process iterates. If $B_* > 0$ we set $s_{i_*} = 1$ and set it to $-1$ otherwise. Then we update $\tilde{f}(\mathbf{s})$ to $\tilde{f}(\mathbf{s} | s_i = s_{i_*})$, add $i_*$ to set $V$, and repeat, conditioning on additional bits until all are selected, yielding estimate for the maximum $\mathbf{s}_*$. In Figures \ref{fig:adv_comp}(a)-(f) we present solver performance using the bias measures of Equations (\ref{eq:bias1}) and (\ref{eq:ell2bias}), which we refer to as bias1 and bias2 respectively throughout the figures. Note that the MeasureBias() algorithm performs conditioning on the strongest biased variables before evaluating subsequent variable bias, meaning that they fix the most strongly biased bits in the approximating function first before proceeding. Empirically, this conditioning eliminates noise by removing half of the search space from consideration at each loop, a process similar to the wavefunction ``collapse" of quantum mechanics. Hard problems cause this process to collapse to the wrong bits, however, requiring a higher fidelity approximation than what is provided by first-order clause indicators.

 In a similar vein, refining an approximation by adding more clause indicators incurs a tradeoff between the complexity of the additional refinement and the accuracy of the resulting approximation, which is generally solution-dependent and therefore unknown. Such a procedure is amenable to a learning-based analysis, but our solver integrates a simpler heuristic approach inspired by existing SAT solvers. We term the algorithm ClauseNeighbors($\mathbf{s}$). Given an unsatisfying configuration $\mathbf{s}$, we first evaluate the set of unsatisfied clauses $S = \{ m | m \in [M],  k_m(\mathbf{s}) = 1\}$. After constructing $S$ we individually flip each variable appearing in any clause in $S$, one at a time, and add any \textit{new} unsatisfying clauses arising from the variable flip to a new set $L$. After flipping all relevant variables, we form the set $U = S \cup L$. Similar constructions arise in, e.g., GSAT \cite{selman1993local}. In the final algorithm, we form all possible second-order products of clause indicators from $U$ to include as new columns of our approximation matrix $\mathbf{A}$. 
 
\begin{algorithm}
 \caption{ClauseNeighbors}
 \begin{algorithmic}
 \renewcommand{\algorithmicrequire}{\textbf{Input:} Configuration $\mathbf{s}_*$}
 \renewcommand{\algorithmicensure}{\textbf{Output:} Set of clause indices (``neighbors") $U$} 
 \REQUIRE in 
 \ENSURE out \\ 
 \textit{Initialization}: Construct $S = \{ m | m \in [M],  k_m(\mathbf{s}_*) = 1\}$ \\ 
 \STATE $U \leftarrow S$
 \FOR{$i \in S$ }
    \FOR{variables $\ell$ in clause $i$}
        \STATE Flip variable $\ell$ in $\mathbf{s}_*$ to form $\mathbf{v}$.
        \STATE Form $L = \{ m | m \in [M], k_m(\mathbf{v}) = 1 \}$
        \STATE $U \leftarrow U + L $
    \ENDFOR 
 \ENDFOR
 \STATE \textbf{return} $U$
 \end{algorithmic}
\end{algorithm}

\begin{figure*}[]
\begin{center}
\includegraphics[width=\textwidth]{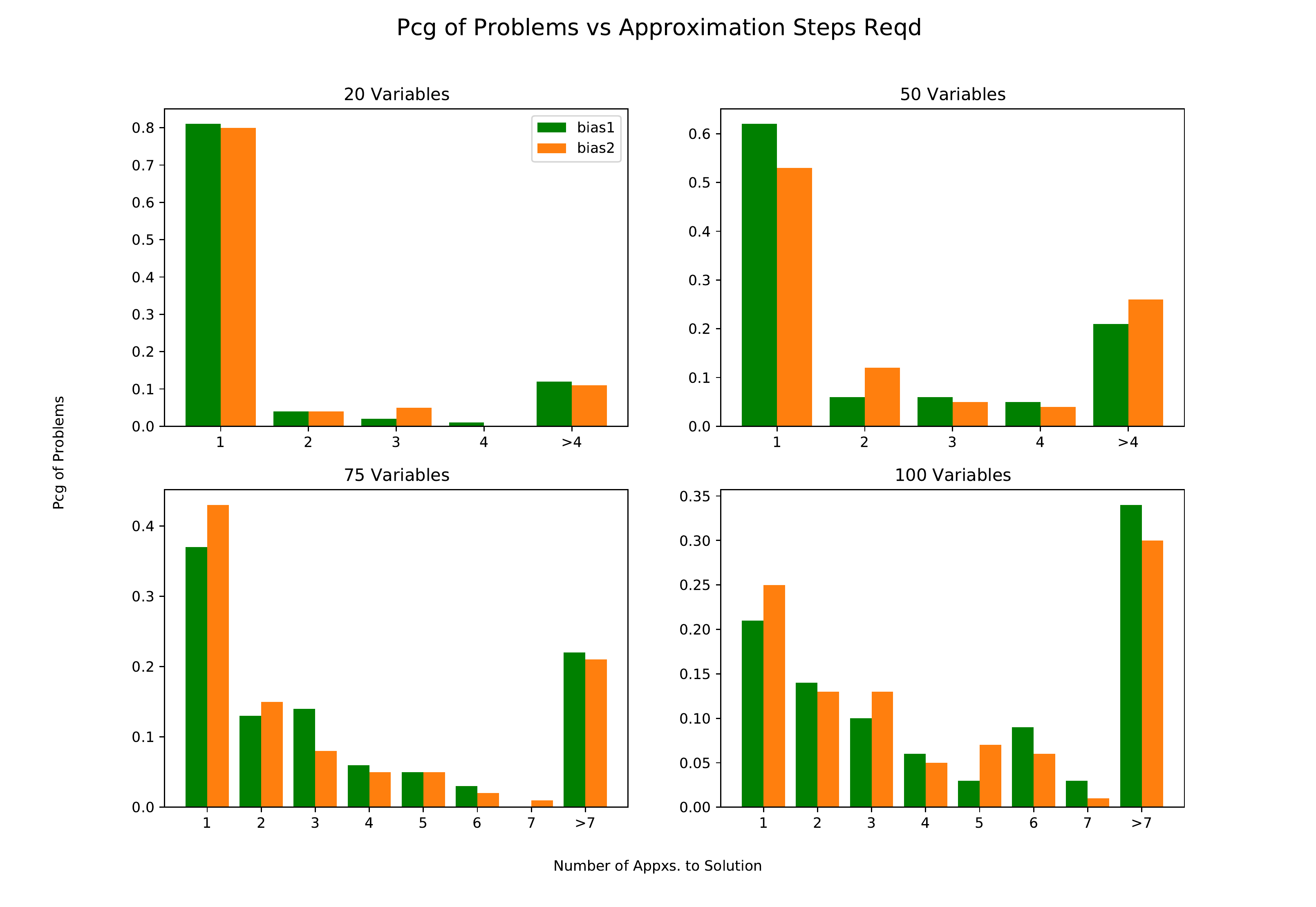}
\end{center}
\caption{The charts depict the number of approximations required before AmplificationSAT returns a satisfying solution. Each chart was generated using 100 satisfiable instances at the given variable size, using the solver method of Algorithm 1. The distributions appear to capture an exponential behavior, with the majority of random instances being solvable within a few approximation refinements, and a growing number of hard instances requiring significantly more refinements as the number of variables increases. For 100 variable problems, a 6000 step annealing linear schedule was used with temperature parameters defined in Section \ref{sec:res_num}}
\label{fig:num_approx}
\end{figure*}
 
In some cases, all second-order pairs are already included in the matrix $\mathbf{A}$ (for example, if a refinement does not change the approximated solution). This represents an opportunity to expand the refinement criteria. Because the heuristic procedure is exhausted in these instances, ClauseNeighbors() instead selects a single clause uniformly randomly and integrates all new second-order pairs involving that clause into the approximation matrix. The process is completely random, as the refinement is not dependent on the properties of the incorrect approximation. However, as demonstrated by Figure \ref{fig:randomness_success}, the addition of random second-order clauses improves the success rate of the solver significantly. Across all problem sizes evaluated, a large proportion of the problems which are not solved using the heuristic strategy are successfully resolved with the addition of random second-order clauses. 
 
\begin{figure*}[]
\begin{center}
\includegraphics[width=\textwidth]{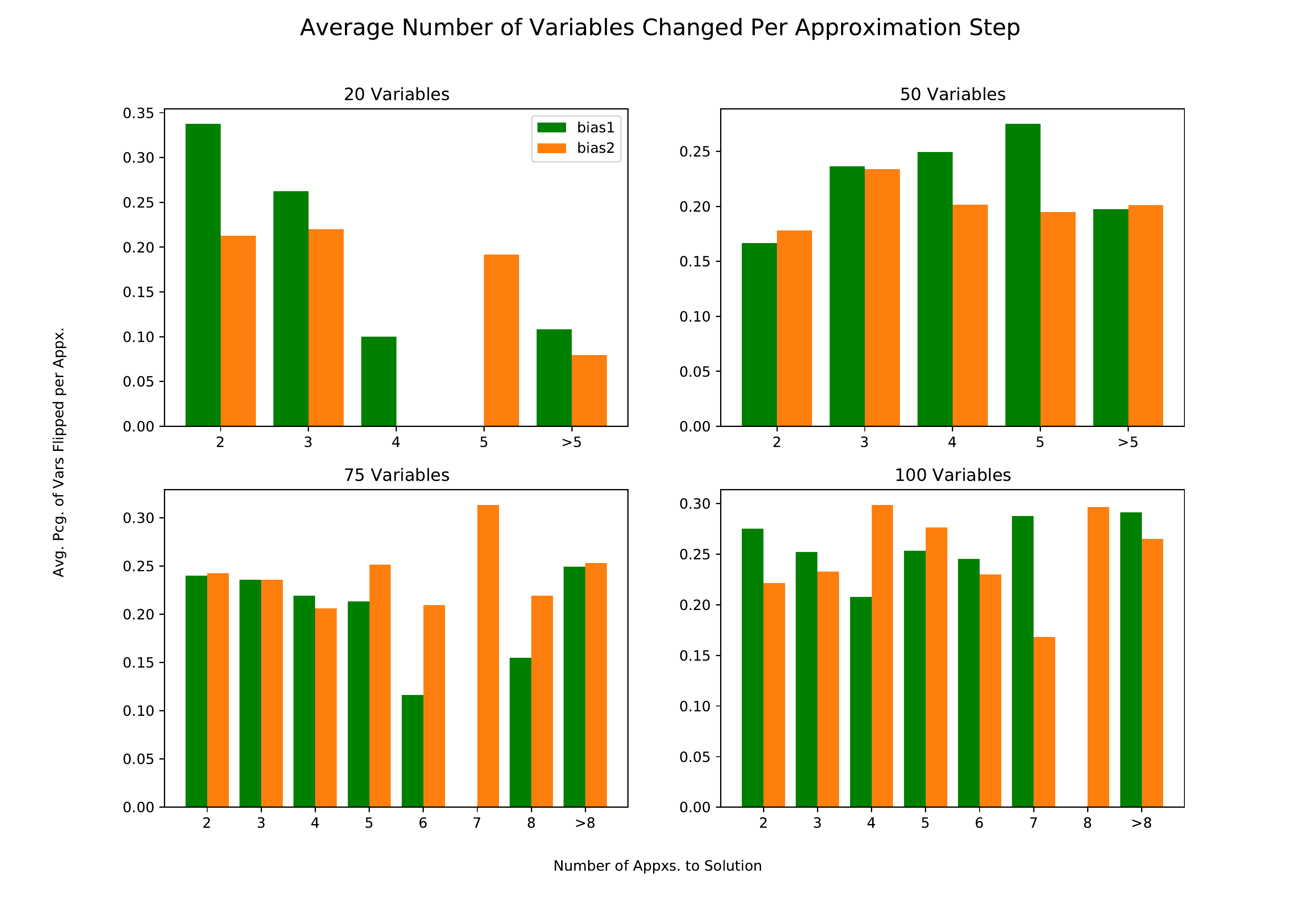}
\end{center}
\caption{The charts capture the average number of variables changed per approximation step versus the number of approximations required to solve the instance. All instances requiring refinement displayed a high average number of variables changed, typically somewhere around one-fourth of the variables changed per refinement. This suggests that AmplificationSAT visits largely disconnected areas of Hamming space with only incremental refinements to the approximation, an advantageous characteristic for pairing with local search methods and problems exhibiting distinct clusters of solutions.}
\label{fig:avg_variables_changed}
\end{figure*}

After each approximation round, the solver performs a local search (e.g. low-temperature annealing). This is a common procedure for global solvers as it allows for exploration of the neighborhood of the approximate solution produced by the solver. Local search after each approximation step ensures that the result of each computation is fully utilized. Analysis of the successive approximations produced by AmplificationSAT demonstrate that the algorithm produces successive candidate solutions that have a large Hamming distance between each other (Figure \ref{fig:avg_variables_changed}). Adding local searches to each refinement, therefore, increases the probability that the search space is effectively explored. The complementary roles of the approximation algorithm and the local search yield a powerful hybrid solver framework that is amenable to a variety of local search procedures in place of simulated annealing.

More advanced heuristic refinements can potentially leverage the ``clustering" of solutions observed in the most difficult random SAT problems to enable the design of approximation algorithms that visit disconnected clusters of solutions \cite{braunstein2005survey}. For example, by logging unsatisfying solutions which are far from each other in Hamming distance, and the associated approximations, such approximations can be refined separately in different manners. This approach stands in stark contrast to local solver paradigms, where solvers traverse the configuration space through weight-1 changes to the input state. For similar reasons, local approaches often perform significantly better with the addition of random restarts or parallelism as this allows them to explore a broader portion of the configuration space; for practical purposes it is often better to start a search process over from a new point than to wait for a local search to exit an attractive local minimum with low-weight updates to the input state. The concept of solution clustering has also been applied successfully to ``inject" local solvers with a type of global awareness of the problem optimization landscape, greatly improving local solver performance \cite{qasem2009learning}. Furthermore, information on correlations between local optima, cost function design, and attraction basins characterizing the optimization landscape may be integrated into the design of global or global-aware solvers \cite{ercsey2011optimization,tayarani2013landscape,lanham2021quantum}.

The AmplificationSAT solver is inherently incomplete, meaning that there is no obvious termination criteria that ensures a problem to be unsatisfiable. For this reason, a fully implemented solver requires a timeout. With a timeout in effect, if $s_{\mathrm{final}}$ is satisfying at any point, the problem is declared SAT, while a non-satisfying $s_{\mathrm{final}}$ at timeout yields UNSAT. The timeout period represents a design parameter that greatly affects the performance of the solver. Figure \ref{fig:adv_comp} demonstrates the difference in performance as a function of timeout for a particular instance of AmplificationSAT, which also applies to the other incomplete solvers in the comparison. 
 
\subsection{Results} 
\label{sec:res_num}
We measured solver performance against FourierSAT \cite{kyrillidis2020fouriersat} and against simulated annealing (SA) with linear temperature schedule in attempt to compare to local approaches and those with similar methodology. We used 100 satisfiable benchmark problems of varying variable and clause sizes. All problems were satisfiable 3-SAT formulas generated uniformly randomly near the hardness threshold and were sourced from the SATLIB library \cite{satlib}. We selected a linear temperature schedule for SA with a dependence on the number of variables. The rule was $T_{\textrm{max}} = 0.1 n$ to $T_{\textrm{min}} = 0.5^{-4}/n$ in a number of steps also dependent on the number of variables (see Figure descriptions). We repeated the annealing algorithm twice per approximation to provide multiple opportunities to avoid local minima. The cost function for SA was adapted from \cite{gu1999optimizing} and simply reflected the number of unsatisfied clauses at a given point. Our FourierSAT implementation followed that of \cite{kyrillidis2020fouriersat} directly. We did not parallelize any algorithms. Due to the significant theoretical differences in the solvers, they were all executed in the same Python computing environment with no particular optimizations to encourage a fair comparison, with the goal being to demonstrate that AmplificationSAT is a relatively competitive solver using novel approximation concepts. 

Figure \ref{fig:adv_comp} highlights two contrasting performance regions for the AmplificationSAT solver. In the short-timeout period, AmplificationSAT excels over local approaches, but the success rate appears to saturate as the timeout periods increase. Local solvers, on the other hand, demonstrate steadily improving success rates with time. AmplificationSAT produces good early estimates for solutions compared to local solvers. The early underperformance of local solvers is partially explained by the ``burn-in" time required for them to produce points that reflect the distributions being estimated \cite{mitra1986convergence}. The AmplificationSAT saturation in performance for longer timeouts likely reflects inadequate heuristic design principles for both the bias implementation MeasureBias() and the approximation refinement principle ClauseNeighbors(), where successive refinements fail to improve the success probability. Performance with long timeout periods is roughly comparable between the solvers, however, as demonstrated by the outperformance multiple. 

Over the longest timeout periods, the local solvers generally outperform. For 20 variable problems at 1 minute per problem, the final success rates were, for AmplificationSAT bias1 93\%, for AmplificationSAT bias2 92\%, for SA 100\%, and for FourierSAT 100\%. For 50 variable problems at 2 minutes per problem the final success rates were, for AmplificationSAT bias1 86\%, for AmplificationSAT bias2 77\%, for SA 89\%, and for FourierSAT 72\%. For 75 variable problems at 3 minutes per problem the final success rates were, for AmplificationSAT bias1 67\%, for AmplificationSAT bias2 57\%, for SA 72\%, and for FourierSAT 20\%.

 The intention of the performance comparison between the bias metrics of Equations (\ref{eq:bias1}) and (\ref{eq:ell2bias}) was to explore a tradeoff between complexity and accuracy, as Equation (\ref{eq:bias1}) biases are constructed using only degree-1 Fourier coefficients and are easy to evaluate, while $\ell_2$ biases use all coefficients of the approximating function. Surprisingly, however, the low-complexity bias measure generally outperformed the $\ell_2$ bias. This behavior can be better understood by interpreting bias1 as a weighted averaging operation which could act to smooth out noisy function values at non-solutions. On the other hand, the $\ell_2$ bias computes a squared sum over function partitions, and so all noisy function values are directly reflected in the sum. This emphasizes that the tradeoff between complexity and accuracy is greatly dependent on proper bias function design. Integrating higher-order coefficients does not straightforwardly improve the solution approximations returned by the solver, but requires a complementary procedure that intelligently uses the additional information. Figure \ref{fig:num_approx} further suggests that the outperformance of AmplificationSAT with bias1 in Figure \ref{fig:adv_comp} with respect to time-to-solution is due to the more efficient evaluation time of the simpler bias function. Approximation functions evaluated with bias2 appear to require somewhat fewer approximation refinements, especially for larger variable sizes, but the biases take longer to evaluate in a manner that offsets the time-to-solution. The additional information leveraged by bias2, therefore, unsuccessfully navigates a tradeoff between approximation accuracy and complexity. 
 
In Figure \ref{fig:avg_variables_changed} we see that the approximation refinements yield candidate solutions in disconnected parts of Hamming space. This is an advantageous feature, especially for the design of hybrid local-global solvers, as it increases the probability that local solvers depart from different attraction basins and arrive at independent feasible solutions. It also suggests a broad exploration of the search space, although the data do not rule out instances where the solution repeatedly toggles between two feasible solutions over additional refinements. The ability of AmplificationSAT to produce multiple unrelated candidate solutions in Hamming space with only a small number of approximation refinements motivates a secondary use as a preliminary generator of initial points for highly parallelizable local solvers, e.g. \cite{kyrillidis2020fouriersat}. 

\subsection{Future Work}
\label{sec:future}
The AmplificationSAT solver achieves competitive performance against local solvers only using $\mathcal{O}(M^2)$ clause indicator functions. The analysis invites further exploration of performance tradeoffs using different measures of bit bias. Performance associated with bias functions can potentially be improved with more advanced decimation strategies, e.g., those involving unit-clause propagation, backtracking, and selection of the order of bit conditioning. More broadly, choices about the pairing of approximation construction and bias metrics relate intimately to questions about what is achievable in an inference setting with bounded computation or low-degree polynomials (see, e.g. \cite{kunisky2022notes,achlioptas2008algorithmic}). We note, additionally, that there is nothing preventing modification of the approximation method  to include high-degree multilinear polynomials as well. The pairing of bias function and approximation refinement path has the potential to transform the structure of the algorithm from one that fundamentally weights biases in proportion to the frequency of a literal appearing in a clause to one that integrates deeper information about the problem, such as higher-order correlations and non-standard partitions of the binary search space. 

The AmplificationSAT framework easily extends to broader constraint satisfaction problems than SAT, so long as the key modularity and factorization properties are present in the problem structure. These properties are also features of mixed constraint satisfiability problems, such as those involving cardinality and XOR. Various constraints including cardinality and XOR are more rigorously developed in the framework of Boolean Fourier analysis in \cite{kyrillidis2020fouriersat}. The demonstrated relevance of an  amplitude-amplification-inspired method in these classical settings may provide opportunities to apply this approach to other computing and optimization problems of fundamental importance. The Boolean constraint indicator functions treated in this work bear a primitive resemblance to amplitude amplification oracles in that they distinguish states (or subspaces) as either ``good" or ``bad" for a particular constraint. This invites further exploration into the connection between tools for manipulating Boolean constraint indicator functions and canonical quantum algorithms that make use of amplitude amplification. 

\section*{Acknowledgment}
The author would like to thank Travis Cuvelier for valuable discussion. This work was supported by an internal research and development grant from Applied Research Laboratories, The University of Texas at Austin.

\bibliographystyle{IEEEtran}
\bibliography{sat_solver_aa.bib}

\end{document}